# Electron–Phonon Coupling Mode in Excitonic Insulator

Akitoshi Nakano[1], Takumi Hasegawa[2], Shinya Tamura[1],
Naoyuki Katayama[1], Satoshi Tsutsui[3], and Hiroshi Sawa[1]

[1] *Department of Applied Physics, Nagoya University, Nagoya 464-8603, Japan*
[2] *Graduate School of integrated Arts and Sciences,
Hiroshima University, Higashi-Hiroshima, Hiroshima 739-8521, Japan*
[3] *Japan Synchrotron Radiation Research Institute (JASRI), SPring-8, Hyogo 679-5198, Japan*

$Ta_2NiSe_5$ is considered a promising excitonic insulator (EI) candidate with slight phonon contributions, since it exhibits a tiny orthorhombic-to-monoclinic structural distortion at 328 K without any superlattice structure. Our synchrotron inelastic x-ray scattering measurements reveal strong electron–optical-phonon coupling occurring at temperatures higher than the transition temperature. Density functional theoretical calculations indicate that two coupled optical modes arise due to the vibration of Ta and Se ions. Further, the two modes are "frozen" such that Ta and Se approach each other, forming atomic-displacement-type electric dipoles in the monoclinic phase. The characteristic of electronic toroidal moment formation by the antiferro-arrangements of electric dipoles is the universality of EI between $Ta_2NiSe_5$ and $1T$-$TiSe_2$.

Excitonic insulators (EIs) have attracted considerable interest as a class of strongly correlated electronic systems different from Mott insulators. In EIs, the spontaneous band hybridization caused by the Coulomb attraction between the orthogonal conduction (electron) and valence (hole) bands opens the bandgap and drives the system into the insulating state [1-6].

In this regard, $1T$-$TiSe_2$ is a potential EI whose a hole pocket of the Se $4p$ orbital and electron pockets of the Ti $3d$ orbital locate at the Γ and L points of the Brillouin Zone (BZ), respectively. As the electron and hole pockets are separated in $q$-space, $1T$-$TiSe_2$ exhibits a commensurate superlattice structure below 200 K, forming electron–hole pairs [7-9]. However, due to the superlattice period, phononic mechanisms other than the excitonic phase, such as the charge-density wave (CDW) and band-type Jahn–Teller effect, have also been proposed [10, 11] to explain the transition, but the transition mechanism is still unclear.

$Ta_2NiSe_5$ has recently been proposed as another EI. It was earlier considered a direct-gap semiconductor with a formal valence state of $Ta^{5+}$ ($5d^0$), $Se^{2-}$($4p^6$), and $Ni^{0+}$($3d^{10}$) [12]; thus, there is no degree of freedom such as charge, orbital, or spin. However, $Ta_2NiSe_5$ exhibits a structural phase transition (orthorhombic to monoclinic) at 328 K [13]. The origin of this transition has remained unclear. Meanwhile, based on subsequent observations of extremely flat valence-band dispersion via angle-resolved photoemission spectroscopy [14], an excitonic scenario has been proposed wherein thermally excited electrons in the Ta $5d$ orbital and holes in the hybridized orbital between Ni $3d$ and Se $4p$ form spin-singlet excitons. This scenario is also supported by theoretical calculations [15-19]. In this regard, recent experimental studies on $Ta_2NiSe_5$ have focused on the purely electronic driven excitonic phase that does not exhibit a superlattice period accompanying CDW [20-22]. In this context, here, we investigate electron–phonon coupling in $Ta_2NiSe_5$.

Figure 1(a) depicts the streak diffuse scattering observed in our x-ray diffraction (XRD) measurements of $Ta_2NiSe_5$ just above the structure phase transition temperature, which strongly suggests that particular phonon modes are involved in EI formation in $Ta_2NiSe_5$. Therefore, to clarify the role of phonons accompanying EI formation in $Ta_2NiSe_5$, we performed inelastic x-ray scattering (IXS) studies. Further, we performed synchrotron XRD measurements to clarify the existence of the "frozen" phonon mode in the low-temperature monoclinic phase.

$Ta_2NiSe_5$ single-crystals were synthesized by chemical vapor transport with $I_2$ as a transport agent [23]. IXS experiments were performed at BL35XU in SPring-8 [24]. A Si (11, 11, 11) backscattering setup, whose energy resolution was 1.5 meV with a wave length of 0.57011 Å (21.747 keV), was chosen. A $Ta_2NiSe_5$ single-crystal (1100 × 220 × 40 μm³) was mounted on a closed-cycle cryostat installed on an IXS spectrometer of BL35XU. The $Q$-resolution was (0.05, 0.05, 0.11) in the reciprocal lattice unit for transverse-wave measurements. Single-crystal XRD experiments were carried out at BL02B1 in SPring-8 with a wavelength of 0.3572 Å [25]. A gas-blowing device was used for sample temperature control. An imaging plate was used to obtain two-dimensional diffraction patterns via oscillation photography methods, and single XRD data were analyzed using the *SHELXL-97* least-squares program [26]. The ABINIT package was employed to optimize crystal structure and calculate phonon modes with the framework of local density approximation density functional perturbation theory (DFPT) [27-34]. The detail of the calculation is de-scribed in the supplement material.

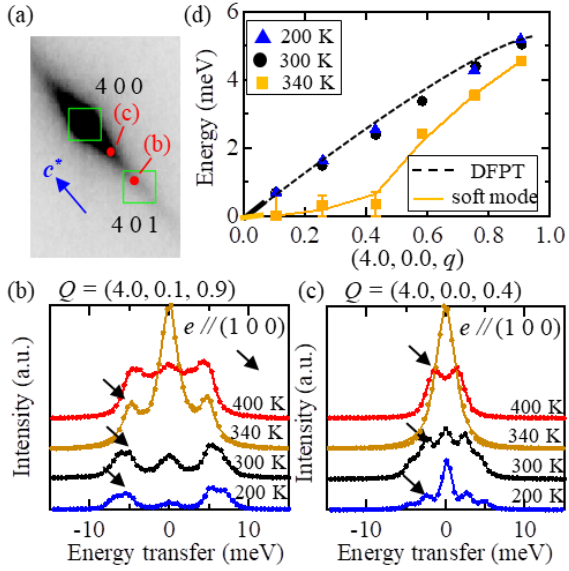

FIG. 1. (a) Streak diffuse scattering in XRD measurements at 330 K. Green squares indicate Bragg peak positions. IXS spectrum at different temperatures at (b) $Q = (4.0, 0.1, 0.9)$ and (c) $Q = (4.0, 0.0, 0.4)$ Black arrows indicate the corresponding TA phonon mode. (d) Energy dispersion of TA mode in Γ-Z-Γ direction.

Figure 1 shows the IXS results of the transverse acoustic (TA) phonons. Since this TA mode is degenerate at the Z point due to space group symmetry, it connects smoothly from $Q = (4.0, 0.0, 0.0)$ (Γ point) to $Q = (4.0, 0.0, 1.0)$ (next Γ point). Figure 1(b) depicts the IXS spectrum at $Q = (4.0, 0.1, 0.9)$, the vicinity of the Γ point of the BZ. At 400 and 340 K, we observe elastic scattering and an inelastic scattering excitation around 6 meV, respectively. Although the inelastic scattering excitation at 8 meV appears below 300 K, this additional excitation is not assigned perfectly. One possible reason is that the folded dispersion at the Γ point may have increased in intensity due to the structural phase transition, consistent with this transition occurring at 328 K. However, the position and linewidth of each inelastic scattering excitation at this momentum do not change significantly between 200 and 400 K.

The IXS spectrum at $Q = (4.0, 0.0, 0.4)$, the vicinity of the Z point of the BZ, shows a distinct temperature dependence (Fig. 1(c)). Although inelastic scattering excitations can be observed around 3 meV at 400 K, they cannot be confirmed at 340 K within the energy resolution range. Below 300 K, the 3-meV inelastic scattering excitations appear again. Further, the elastic scattering intensity increases with temperature decrease from 400 to 340 K. Subsequently, the elastic scattering intensity decreases rapidly at 300 K. This behavior is consistent with the temperature dependence of diffuse scattering (Fig. S1).

The TA phonon dispersion in the Γ-Z-Γ direction is shown in Fig. 1(d). At 200 and 300 K, the dispersion increases from zero at the Γ point; therefore, this phonon behaves as a typical TA-mode phonon and is assigned to $B_{2g}$ symmetry via DFPT calculations. However, at 340 K, there is strong TA-mode softening, wherein the energy decreases to nearly zero in the momentum range up to $q \sim 0.4$. This implies that the interatomic interaction vanishes within 1 or 2 unit cells along the $c$-axis. There are few reports of softening over such a wide wavenumber range, and only an example of ferroelectric LAT [36]. We note that diffuse scattering or the significant soft mode was not observed in the Γ-X direction within the resolution range (not shown). Since the acoustic phonon at $q \sim 0$ exhibits strong softening, the $Ta_2NiSe_5$ structural phase transition is concluded to be an elastic phase transition, with strain as an order parameter. Although $Ta_2NiSe_5$ has a layered crystal structure with each layer composed of Ta-Se octahedrons and Ni-Se tetrahedrons [23], the "strain" herein is the relative shift of the quasi-one dimensional (1D) structure of the transition metal "chained" along the $a$-axis direction. Indeed, the $\beta$ angle of the unit cell corresponds to the order parameter of ferroelastic transition. On the other hand, optical phonon modes, which generally cause electric fields strongly, seem to be important to interact with electronic state since excitons are formed by electrons and holes.

The transverse optical (TO) modes, which also exhibit anomalous behaviors, are shown in Fig. 2. Figure 2(a) shows the IXS spectrum at 300 K measured at $Q = (4.0, 0.0, 1.85)$. In addition to the elastic scattering, two inelastic scattering excitations are observed in the vicinity of 8 and 11 meV. In contrast, in the 400 K spectrum, the spectral linewidth significantly increases. Further, the excitation corresponding to 11 meV observed at 300 K cannot be confirmed; a shoulder is present in the vicinity of 4 meV instead.

The IXS spectrum at 300 K and the DFPT calculation exhibits good agreement (Fig. S2). The two TO modes in Fig. 2(a) are assigned to $B_{2g}$ symmetry, corresponding to atomic vibrations, as shown in Figs. 2(c) and (d). In the mode of 8 meV (11 meV), Ta (Se) ions mainly oscillate along the 1D chain structure direction ($a$-axis) as an anti-phase vibration between 1D chains. Hereafter, we describe these modes as $B_{2g}^{Ta}$ and $B_{2g}^{Se}$. There is one more optical mode (with $B_{3u}$ symmetry) overlapping with $B_{2g}^{Se}$; this is described in the supporting information. In contrast, the IXS spectrum at 400 K could not be reproduced by DFPT calculations, possibly as an effect of electronic fluctuations not complemented by the DFPT calculation. Thus, we performed Lorentzian profile-fitting of the IXS spectrum at 400 K assuming that two excitations appear in the energy range up to 15 meV, as at 300 K.

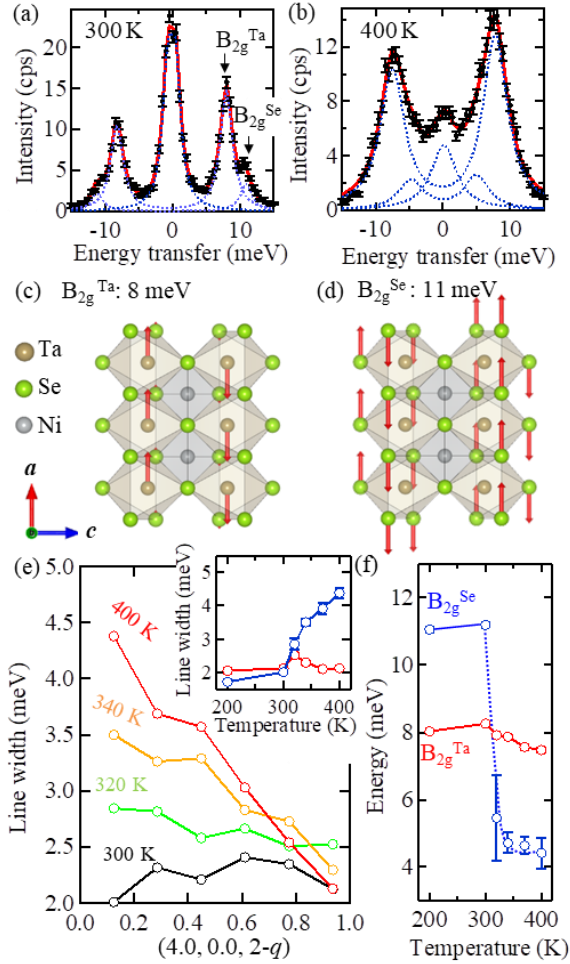

FIG. 2. IXS spectrum at $Q = (4.0, 0, 1.85)$ at (a) 300 K and (b) 400 K. (c) and (d) Schematic of collective vibration of $B_{2g}^{Ta}$ and $B_{2g}^{Se}$ modes, respectively. (e) Linewidth dispersion in the $\Gamma$-Z-$\Gamma$ direction at various temperatures. Inset shows the temperature dependences of the linewidth at $Q = (4.0, 0.0, 1.85)$ (blue) and $Q = (4.0, 0.0, 1.06)$ (red). (f) Temperature dependence of energy of corresponding $B_{2g}$ modes.

Figure 2(e) plots the linewidth of the $B_{2g}^{Ta}$ mode at various temperatures. At 400 K, the linewidth apparently increases towards $Q = (4.0, 0.0, 1.85)$ with an increasing ratio of 200%. With temperature decrease, such anomalies are suppressed, and there is almost no $q$ dependency at 300 K. The inset of Fig. 2(e) plots the linewidth at $q = 0.15$ and $q = 0.9$ against temperature. The anomaly near the $\Gamma$ point is suppressed from 400 K towards T ~ 320 K, which corresponds to the structural phase transition temperature. Therefore, this abnormality appears to be a precursor to the phase transition.

In Fig. 2(f), the positions of the two excitations obtained as a fitting of the IXS spectrum at $q = 0.15$ are plotted against temperature. The almost-constant excitation at 8 meV is assigned as $B_{2g}^{Ta}$ mode, and the second is considered as $B_{2g}^{Se}$ mode. In the graph, the other excitation has a low energy of ~4 meV on the high-temperature side, suggesting that $B_{2g}^{Se}$ mode is softened just above the phase transition. The energy dispersions of the $B_{2g}$ modes are shown in Fig. S3.

We next review the crystal structure of the low-temperature monoclinic phase based on the IXS findings. The atomic displacement determined by the crystal structure analysis conducted at 30 K is shown in Fig. 3(a). The orange arrows indicate the magnitude and direction of the atomic displacements. This distortion can be interpreted as a combination of $B_{2g}^{Ta}$ and $B_{2g}^{Se}$ modes shown in Figs. 2(c), (d). We note that $B_{2g}^{Ta}$ and $B_{2g}^{Se}$ are frozen; thus, Ta and Se approach each other, which is obviously different from a Peierls transition or Jahn–Teller distortion.

On the other hand, Ni and Se(1) constituting the 1D chain locate at special positions with two-folded rotation symmetry, and their fractional coordinates hardly change through the structural phase transition. Thus, we conclude that the optical modes $B_{2g}^{Ta}$ and $B_{2g}^{Se}$ drive the structural phase transition, and the amplitude when the optical mode is frozen is the hidden order parameter of the phase transition (Fig. S3). This is also supported by the fact that the freezing amplitude of the $B_{2g}$ modes is also suppressed when the excitonic phase transition is suppressed by applying hydrostatic pressure [37] (Table S1).

It is noteworthy that although the TA phonon mode shows significant softening at the transition temperature, the structural phase transition is described as freezing of the $B_{2g}$ modes. Although we have clarified that the TO phonon mode drives the Ta$_2$NiSe$_5$ phase transition, actually it is common with another EI candidate, 1$T$-TiSe$_2$.

In 1$T$-TiSe$_2$, the TO phonon reportedly becomes soft at the L point in the momentum space just above the phase transition with increase in the phonon linewidth toward the L point [38]. This linewidth increase of the optical phonon is caused by strong electron–phonon coupling. Similarly, the phase transition is also possibly driven by strong electron–phonon interaction in Ta$_2$NiSe$_5$. In contrast to 1$T$-TiSe$_2$, the linewidth increase at the $\Gamma$ point in Ta$_2$NiSe$_5$ (with its direct-bandgap feature) reflects the strong influence of the Fermi surface at the $\Gamma$ point. This is consistent with the proposed excitonic theory as per which Coulomb attraction between electrons and holes occurs at the $\Gamma$ point [15].

We note that the vibration patterns of $B_{2g}^{Ta}$ and $B_{2g}^{Se}$ in Ta$_2$NiSe$_5$ are similar to each of the three $A_u$ modes (Triple-$Q$) in 1$T$-TiSe$_2$ [39]. Further, it is even common that Ti$^{4+}$ (with its $d^0$ configuration) and Se freeze to approach each other [9]. These universalities strongly suggest that the physical phenomena occurring in both materials are the same.

Moreover, $Ta_2NiSe_5$ is different in that the $B_{2g}^{Ta}$ and $B_{2g}^{Se}$ modes have the same symmetry as that of the TA phonon mode; thus, they can be hybridized through electronic fluctuations. Consequently, anomalous softening of the TA phonon may occur, as in Fig. 1(d).

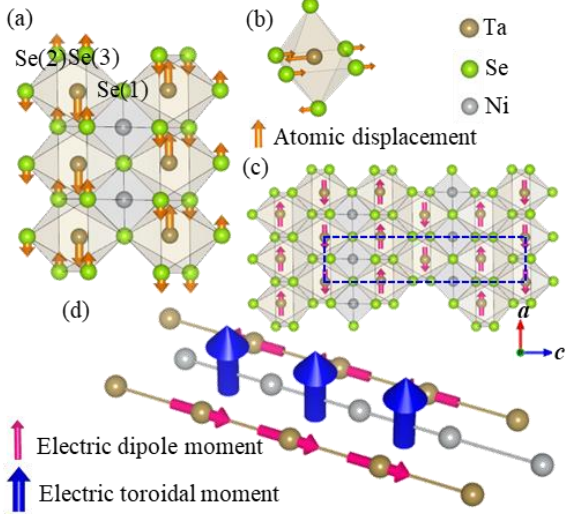

FIG. 3 (a) Atomic displacement in monoclinic phase as per XRD analysis. (b) Local structure of Ta-Se octahedron. (c) Antiferroelectric arrangement of electric dipole moment of Ta-Se octahedrons. (d) Formation of electric toroidal moment.

To realize the excitonic phase, the electronic state of Se appears particularly important, since the softening of the $B_{2g}^{Se}$ mode occurs just before the structural phase transition in $Ta_2NiSe_5$. This Se electronic state is plausibly explained by the existence of the ligand hole previously reported via x-ray photoemission spectroscopy measurements [14]. The unusual $3d^9$ valence state of Ni may play an important role in the formation of the ligand hole [14]. From this perspective, the two frozen $B_{2g}$ modes in $Ta_2NiSe_5$ can be regarded as a bond order between Se with a slight ligand-hole and Ta with a slight electron, as in $1T$-$TiSe_2$ [40]. Furthermore, $Ta_2NiS_5$ and $TiS_2$, which are isostructural with $Ta_2NiSe_5$ and $TiSe_2$, respectively, do not exhibit the excitonic phase transition [9, 13]. This suggests that exchange from Se to S affects both the electronic bandgap size and electron–phonon interactions since the light mass of S than Se increases the energy of the $B_{2g}$ mode.

Meanwhile, many universal physical properties have been reported for both $Ta_2NiSe_5$ and $1T$-$TiSe_2$, such as anomalous band dispersion flattening [7, 14] and similar pressure-temperature electronic phase diagrams [41, 42]. Considering these findings on electron–phonon coupling and structural distortion, we may appropriately categorize $Ta_2NiSe_5$ and $1T$-$TiSe_2$ under a new framework: the cooperative phenomena between electron–hole (excitonic) and electron–phonon interactions in systems consisting of $d^0$ cations and Se. In other words, our findings can shed new light on the controversial phase transition of $1T$-$TiSe_2$.

Finally, in $Ta_2NiSe_5$, we remark that the structural phase transition can be regarded as a certain multipole ordering. Focusing on the local Ta-Se octahedron in Fig. 3(b), we note that an atomic-displacement-type electric dipole is formed. Figure 3(c) shows the arrangement of such dipole moments as being in the Ta site, indicated by purple arrows. Electric dipoles are aligned in the same direction along the $a$-axis. On the other hand, along the $c$-axis, since they are aligned in the opposite direction, the whole arrangement can be regarded as antiferroelectric. In the case of magnetic dipole moment, it is posited that the antiferromagnetic arrangement can considered as a toroidal moment [43]. Considering this analogy, the antiferroelectric arrangement can be also regarded as an electric (axial) toroidal moment. In the case of $Ta_2NiSe_5$, differing from general antiferroelectric materials, the antiferroelectric distortion occurs with the $q = 0$ structural phase transition; therefore, it can be regarded as ferrotoroidic.

Surprisingly, there is universality between the formation of the antiferroelectric nature of this electric dipole and the toroidal moment between $Ta_2NiSe_5$ and $1T$-$TiSe_2$. In the case of $1T$-$TiSe_2$, an antiferroelectric toroidal ordering is formed in the excitonic phase, in contrast to $Ta_2NiSe_5$ [40]. The relationship between electric toroidal ordering and EI state has already been examined within the simple two-band model [44, 45]. Future studies can focus on the relationship between excitonic order and structural phase transition discussed here via constructing an effective model of $Ta_2NiSe_5$.

In summary, we performed inelastic x-ray scattering measurements of $Ta_2NiSe_5$, which is a highly probable excitonic insulator (EI) candidate. The transverse acoustic phonon exhibited anomalous softening just above the structural phase transition at 328 K, and at higher temperatures, the transverse optical phonon exhibited energy linewidth broadening and softening. The broadened linewidth indicates that strong electron–phonon coupling drives the structural phase transition, as in $1T$-$TiSe_2$, which is another EI candidate. Further, we determined that the atomic displacement in the low-temperature phase is similar between $Ta_2NiSe_5$ and $1T$-$TiSe_2$, in which the $d^0$ cation and Se approach each other. Our findings can provide further insights into the excitonic phase transitions of both $Ta_2NiSe_5$ and $1T$-$TiSe_2$.

We thank Dr. T. Kaneko and Professor. Y. Motome for helpful discussion. Single crystal IXS and XRD meas-